\documentstyle [multicol,aps,prl,graphicx] {revtex} 
\begin{document}
\pagestyle{empty} \draft
\title{The effect of surface roughness on the adhesion of elastic solids}
\author{B.N.J. Persson$^{1,2,\ast}$ and E. Tosatti$^{2,3,4, \dag}$}
\address{$^1$IFF, FZ-J\"ulich, 52425 J\"ulich, Germany}
\address{$^2$International School for Advanced Studies (SISSA), 
Via Beirut 2-4, I-34014, Trieste, Italy}
\address{$^3$INFM, Unita' SISSA, Trieste, Italy}
\address{$^4$ International Centre for Theoretical Physics (ICTP)
P.O. Box 586, I-34014, Trieste, Italy}
\address{$\ast$ e-mail B.Persson@fz-juelich.de}
\address{\dag\ e-mail tosatti@sissa.it}
\maketitle

\begin{abstract}
We
study the influence of
surface roughness on the adhesion of elastic solids.
Most real surfaces have roughness on many different length scales,
and this fact is
taken into account in our analysis. We consider in detail the
case when the surface roughness can be described as a self
affine fractal, and show that
when the fractal dimension $D_{\rm f}
>2.5$,
the adhesion force may vanish,
or be at least strongly reduced.
We consider the block-substrate pull-off force as a function
of roughness, and find a partial detachment transition preceding
a full detachment one.
The theory is in good qualitative agreement with experimental data.
\vskip 0.5cm \noindent 81.40.Pq, 62.20-x \vskip
0.5cm
\end{abstract}

\begin{multicols}{2}
\narrowtext

{\bf 1. Introduction}

Even a highly polished surface has surface roughness on
many different length scales.
When two bodies with
nominally flat surfaces are brought into contact,
the area of real contact will usually
only be a small fraction of the nominal contact area.
We can visualize the contact regions as small areas where asperities from
one solid are squeezed against asperities of the other solid;
depending on the conditions the
asperities may deform elastically or plastically.

How large is the area of {\it real} contact
between a solid block and the substrate? This
fundamental question has extremely
important practical implications. For example,
it determines the contact resistivity and the heat transfer between the
solids. It is also of direct
importance for sliding friction\cite{[1]}, e.g., the rubber friction
between a tire and a road surface, and it has a major influence on the adhesive
force between
two solids blocks in direct contact.
One of us has developed a theory of contact mechanics\cite{[2]},
valid for randomly rough (e.g., self affine fractal)
surfaces, but neglecting adhesion. Adhesion is particular important for
elastically soft solids, e.g., rubber or gelatine, where it may
pull the two solids in direct contact over the whole
nominal contact area.

In this paper we discuss adhesion for randomly rough surfaces.
We first calculate the block-substrate
pull-off force under the assumption that there is complete contact at the
in the nominal contact area. We
assume that the substrate surface
has roughness on many different length scales, and consider in detail the case
where the surfaces are self affine fractal. We also study pull-off
when only partial contact occurs in the nominal contact area.

The influence of surface roughness on the
adhesion between rubber (or any other elastic solid)
and a hard substrates has been studied in a classic
paper by Fuller and Tabor\cite{[3]}. They found that already a relative
small surface roughness can completely remove the adhesion.
In order to understand the experimental data they developed a very simple model
based on the assumption of surface roughness on a single length scale.
In this model the rough surface is modeled by asperities all
of the same radius of curvature and with heights
following a Gaussian distribution.
The overall contact force was obtained by applying the contact theory of
Johnson, Kendall and Roberts\cite{[4]} to each individual asperity.
The theory predicts that the pull-off force, expressed as a fraction of the
maximum
value, depends upon a single parameter, which may be regarded as
representing the statistically averaged competition between the
compressive forces exerted by the higher asperities trying to
prize the surfaces apart and the adhesive forces between the lower
asperities trying to hold the surfaces together. We believe that this
picture of adhesion developed by Tabor and Fuller would be correct {\it if}
the surfaces had roughness on a single length scale as assumed
in their study. However, when roughness occurs on many different length scales,
a qualitatively new picture emerges (see below),
where, e.g., the adhesion force may even vanish
(or at least be strongly reduced), if
the rough surface can be described as a self affine fractal with
fractal dimension $D_{\rm f}
>2.5$. We also note that the formalism used by Fuller and Tabor is
only valid at ``high'' surface roughness, where the area of real
contact (and the adhesion force) is very small.
The present theory, on the other hand, is
particularly accurate for ``small'' surface roughness,
where the area of real contact
equals the nominal contact area.

\vskip 0.5cm

{\bf 2. Qualitative discussion}

\begin{figure}[htb]
\begin{center}
\begin{minipage}{\textwidth}
   \includegraphics[width=0.475\textwidth]{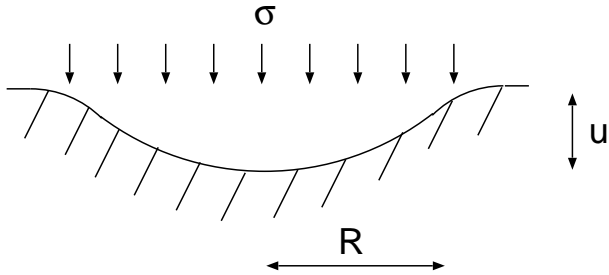}
\end{minipage} \\
\end{center}
\caption{
A uniform stress $\sigma$, acting within a circular area (radius $R$)
on the surface of a semi-infinite elastic medium,
gives
rise to a displacement $u$.
 }
\label{PT1}
\end{figure}

Assume that a uniform stress $\sigma$ acts within a circular area
(radius $R$) centered at a point P on the surface
of a semi-infinite elastic body with elastic modulus $E$, see Fig.\ \ref{PT1}.
This will give rise to a perpendicular displacement $u$ of P by a distance
which is easy to calculate using continuum mechanics:
$u/R \approx \sigma /E$.
This result can also be derived from simple
dimensional arguments. First, note that $u$ must be proportional
to $\sigma$ since
the displacement field is linearly related to the stress field. However, the
only other quantity in the problem with the same dimension as the stress
$\sigma $ is the elastic
modulus $E$ so $u$ must be proportional to $\sigma /E$. Since $R$ is
in turn the only quantity with the dimension of length we get at
once $u \sim (\sigma /E)R$.  Thus,
if 
$h$ and $\lambda$ represent perpendicular and parallel roughness
length scales respectively, then 
if $h/\lambda \approx  \sigma/E$,
the perpendicular pressure $\sigma$ will
be just large enough to deform the rubber to make contact with the substrate
everywhere.
Let us now consider the role of the rubber-substrate adhesion interaction.
When the rubber deforms and fills out a surface cavity of the substrate,
an elastic energy $U_{\rm el}\approx E \lambda h^2$
will be stored in the rubber. Now, if this
elastic energy is smaller than the gain in adhesion energy $U_{\rm
ad}\approx - \Delta \gamma \lambda^2$, 
where $-\Delta \gamma $ is the local change of surface free energy
upon contact
due to
the rubber-substrate interaction (which usually
is mainly of the van der Waals type),
then (even in the absence of the load $F_{\rm N}$) the rubber will deform
{\it spontaneously} to fill out the substrate cavities. The condition $U_{\rm
el}
=-U_{\rm ad}$ gives $h/\lambda \approx
(\Delta \gamma /E\lambda)^{1/2}$. For example, for very rough surfaces
with $h/\lambda \approx 1$, and with parameters typical of rubber
$E=1 \ {\rm MPa}$ and $\Delta \gamma = 3 \ {\rm meV/\AA}^2$, the
adhesion interaction will be able to deform the rubber and completely fill
out the cavities if $\lambda < 0.1 \ {\rm \mu m}$. For very smooth surfaces
$h/\lambda \sim 0.01$ or smaller, so that the rubber will be able
to follow the surface roughness profile up to the length scale
$\lambda \sim 1 \ {\rm mm}$ or longer.

The discussion above assumes
roughness on a single length scale $\lambda$.
But the surfaces or real solids have roughness
on a wide distribution of length scales.
Assume, for example, a self affine fractal surface.
In this case the statistical properties of the surface
are invariant under the transformation
$${\bf x} \rightarrow {\bf x} \ \zeta, \ \ \ \ \ \ \
z \rightarrow z  \ \zeta^H$$
where ${\bf x} = (x,y)$ is the 2D position vector in the surface plane,
and where $0<H<1$. This implies that if $h_0$ is the
amplitude of the surface roughness
on the length scale $\lambda_0$, then
the amplitude $h$ of the surface roughness on
the length scale $\lambda$ will be of order
$$h\approx h_0 \ (\lambda/\lambda_0)^H$$
Thus, the condition $E_{\rm ad} > E_{\rm el}$, i.e.,
$\Delta \gamma \lambda > E h^2$, gives
$$\Delta \gamma \lambda > E h_0^2\left ({\lambda \over \lambda_0}\right
)^{2H}$$ or
$$\left ({\lambda\over \lambda_0}\right )^{2H-1}
< {\Delta \gamma \lambda_0 \over E h_0^2}$$
Hence for $H>1/2$, if $\Delta \gamma \lambda_0/Eh_0^2 \geq 1$
the adhesion will be  important {\it on
all length scales, and complete contact will
occur at the interface}. When $H<1/2$ it is clear that without a
short-distance cut off, {\it adhesion and the area of real contact
will vanish}. In reality, a finite short-distance cut off will always occur,
but this case requires a more detailed study (see
in Sec. 3). Also, in the analysis above we have neglected that the
area of real contact depends on $h$ (i.e., it is of order
$\lambda^2$ only when $h/\lambda << 1$). A more accurate
analysis follows below.

\vskip 0.5cm

{\bf 3. Interfacial elastic and adhesion energies for rough surfaces}

\begin{figure}[htb]
\begin{center}
\begin{minipage}{\textwidth}
   \includegraphics[width=0.475\textwidth]{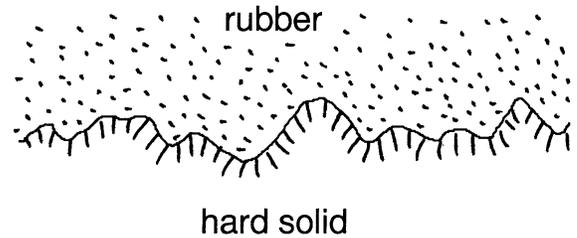}
\end{minipage} \\
\end{center}
\caption{
The adhesion interaction pull the rubber into complete contact
with the rough substrate surface.
 }
\label{PT2}
\end{figure}

Assume that a flat rubber surface is
in contact with the rough surface of a hard
solid. Assume that because of the rubber-substrate adhesion interaction, the
rubber deforms elastically and makes contact
with the substrate everywhere, see Fig.\ \ref{PT2}.

Let us calculate
the difference in free energy between the rubber block in contact
with the substrate and the non-contact case. Let $z=h({\bf x})$ denote
the height of the rough surface above a flat reference plane
(chosen so that $\langle h \rangle = 0$).
Assume first that the rubber is in direct contact
with the substrate over the whole nominal contact area. The surface adhesion
energy is assumed proportional to the contact area so that
$$U_{\rm ad} = -\Delta \gamma \int d^2x \ \left [1+\left (\nabla h({\bf x})
\right )^2\right ]^{1/2}$$
$$\approx -\Delta \gamma \left [A_0+{1\over 2} \int d^2x \
\left (\nabla h \right )^2\right ]\eqno(1)$$
where we have assumed $\mid \nabla h \mid << 1$. Now,
using
$$h({\bf x}) = \int d^2q  \ h({\bf q}) {\rm e}^{i{\bf q}\cdot {\bf x}}$$
we get
$$
\int d^2x \
\left (\nabla h \right )^2$$
$$ = \int d^2x \int d^2q \ d^2q' \ (-{\bf q}\cdot {\bf q'})
\langle h({\bf q})
h({\bf q'}) \rangle \ {\rm e}^{i({\bf q}+{\bf q'})\cdot {\bf x}} $$
$$=(2\pi )^2 \int d^2q \ q^2
\langle h({\bf q})
h({\bf -q}) \rangle $$
$$= A_0 \int d^2q \ q^2 C(q)\eqno(2)$$
where the surface roughness power spectrum is
$$C(q) = {1\over (2\pi )^2}\int d^2x \ \langle h({\bf x})
h({\bf 0})\rangle {\rm e}^{-i{\bf q}
\cdot {\bf x}},\eqno(3)$$
where $\langle...\rangle $ stands for ensemble average.
Thus, using (1) and (2):
$$U_{\rm ad} \approx - A_0 \Delta \gamma
\left [1+{1\over 2} \int d^2q \ q^2 C(q) \right ]\eqno(4)$$

Next, let us calculate the elastic energy stored
in the deformation field in the
vicinity of the interface. Let $u_z({\bf x})$ be the normal displacement field
of the surface of the elastic solid. We get
$$U_{\rm el} \approx -{1\over 2} \int d^2x \ \langle u_z({\bf x})\sigma_z({\bf
x})\rangle$$
$$=-{(2\pi )^2 \over 2} \int d^2q \
\langle u_z({\bf q}) \sigma_z (-{\bf q})\rangle\eqno(5)$$
Next, we know that\cite{[5]}
$$u_z({\bf q})= M_{zz} ({\bf q}) \sigma_z ({\bf q})\eqno(6)$$
where
$$M_{zz} ({\bf q}) = -{2(1-\nu^2)\over Eq},\eqno(7)$$
$E$ being the elastic modulus and $\nu$ the Poisson ratio.
If we assume that complete contact occurs between the
solids, then $u_z=h({\bf x})$ and
from (3) and (5)-(7),
$$U_{\rm el} \approx -{(2\pi )^2 \over 2} \int d^2q \ \langle u_z({\bf q}) u_z
(-{\bf q})\rangle
\left [M_{zz}(-{\bf q})\right ]^{-1}$$
$$={A_0E\over 4(1-\nu^2)}\int d^2q \ q C(q)\eqno(8)$$

The change in the free energy
when the rubber block moves in contact with the substrate is
given by the sum of (4) and (8):
$$U_{\rm el}+U_{\rm ad} =
-\Delta \gamma_{\rm eff} A_0$$
where
$$\Delta \gamma_{\rm eff} =
\Delta \gamma \left [1+\pi \int_{q_0}^{q_1} dq \ q^3 C(q) \right.$$
$$\left. -{\pi E\over 2(1-\nu^2)\Delta \gamma}
\int_{q_0}^{q_1} dq \ q^2 C(q)\right ]
\eqno(9)$$

The theory above is valid for surfaces with arbitrary random roughness,
but will now be applied to self-affine fractal surfaces.
It has been found that many ``natural'' surfaces, e.g., surfaces of
many materials generated by
fracture, can be approximately described as self-affine
surfaces over a rather wide roughness size region. A self-affine
fractal surface has the property that if we make a scale change that is
appropriately different along the two directions, parallel and perpendicular,
then the surface does not change its morphology\cite{[6]}.
Recent studies have shown that
even asphalt road tracks (of interest for rubber friction)
are (approximately) self-affine fractal, with an upper cut-off
length $\lambda_0 = 2\pi /q_0$ of order a few mm\cite{[7]}.
For a self affine fractal
surface\cite{[6],BL}:
$C(q)=0$ for $q<q_0$, while for $q>q_0$:
$$C(q) = {H\over 2 \pi} \left ({h_0\over q_0}\right )^2 \left
({q\over q_0}\right )^{-2(H+1)},\eqno(10)$$
where $H=3-D_{\rm f}$ (where the fractal dimension $2 < D_{\rm f} < 3$), and
where $q_0$ is the lower cut-off wavevector, and $h_0$ is determined by the rms
roughness amplitude, $\langle h^2\rangle = h_0^2/2$.

\begin{figure}[htb]
\begin{center}
\begin{minipage}{\textwidth}
   \includegraphics[width=0.475\textwidth]{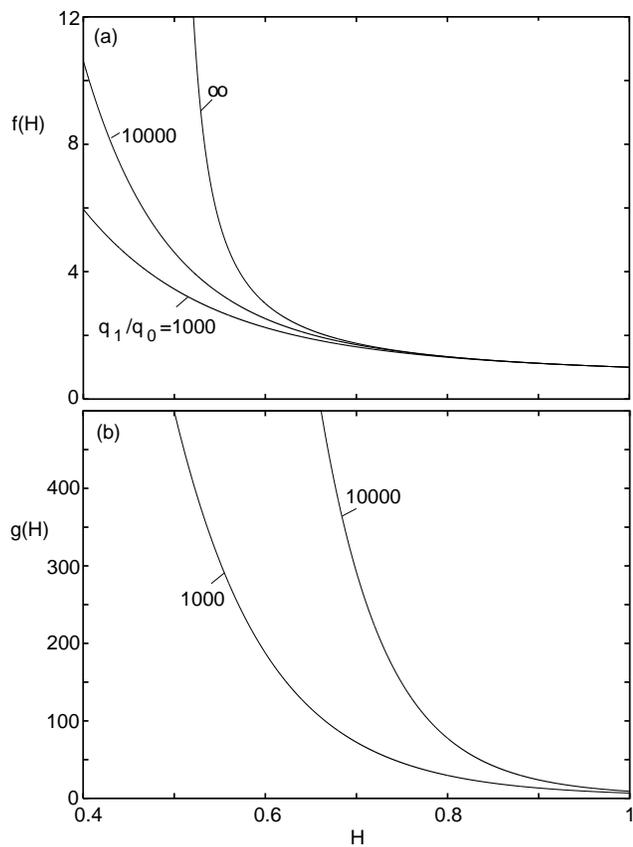}
\end{minipage} \\
\end{center}
\caption{
The functions $f(H)$ and $g(H)$ are defined in the text.
 }
\label{PT3}
\end{figure}

Substituting (10) in (9) gives
$$\Delta \gamma_{\rm eff} =
\Delta \gamma \left [ 1
+{1\over 2} (q_0h_0)^2 g(H)
-{E h_0^2 q_0  \over 4(1-\nu^2)\Delta \gamma}
f(H) \right ]\eqno(11)$$
where
$$f(H)=
{H \over 1-2H}
\left [ \left ({q_1\over q_0}\right )^{1-2H}-1 \right ]\eqno(12)$$
$$g(H)=
{H \over 2(1-H)}
\left [ \left ({q_1 \over q_0} \right )^{2(1-H)}-1 \right ]\eqno(13)$$
If we introduce the length $\delta = 4(1-\nu^2) \Delta \gamma /E$, then
(11) takes the form
$$\Delta \gamma_{\rm eff} =
\Delta \gamma \left [ 1
+ (q_0h_0)^2 \left ( {1\over 2}g(H)
-{1  \over q_0 \delta}
f(H)\right ) \right ]\eqno(14)$$
In Fig.\ \ref{PT3}  we show $f(H)$ and $g(H)$ as a function of $H$. Note that
the
present theory is valid only if $(q_0 h_0)^2 g(H)/2 < 1$, otherwise the
expansion
of the square-root function in (1) is invalid.

Consider first an elastically very soft solid, e.g., jelly.
In this case, using $E\approx 10^4 \ {\rm Pa}$ and
$\Delta \gamma \approx 3 \ {\rm meV/\AA^2}$, we get
$\delta \approx 10 \ {\rm \mu m}$,
and since typically $q_0 = 2\pi /\lambda_0 \sim (10 \ {\rm \mu m})^{-1}$
and $g(H)>>f(H)$, we expect $\Delta \gamma_{\rm eff}
> \Delta \gamma $. Thus, for an (elastically)
very soft solid the adhesion force may increase
upon roughening the substrate surface. This effect has
been observed experimentally\cite{[8]}, and the present theory
explains under exactly what conditions that will occur. 

For most ``normal'' solids,
$\Delta \gamma \approx Ea$, where $a$ is an atomic distance (of order
$\sim 1 \ {\rm \AA}$) and $E$ the elastic modulus. Thus,
$\delta \sim a \sim 1 \ {\rm \AA}$ and
typically $1/q_0\delta \sim 10^4$ so that the
(repulsive)
energy stored in the
elastic deformation field in the solids at the interface,
and proportional to $f(H)$, largely overcomes
the
increase in adhesion energy derived from the
roughness induced increase in the contact area,
described by the term $(q_0h_0)^2g(H)/2$.

Let us note the following very
important fact. Many solids respond in an elastic manner when exposed
to rapid deformations, but flow plastically on long enough time
scales. This is
clearly the case for un-cross-linked glassy polymers, but it is also
to some extent the case
for rubbers with cross-links. The latter materials behave as relative hard
solids when exposed to high-frequency perturbations, while they deform as
soft solids when exposed to low-frequency perturbations.
Thus, when such a solid is squeezed rapidly
against a substrate with roughness
on many different length scales,
a large amount of elastic energy
may initially be stored in the local (asperity induced)
deformation field at the interface.
However, if the
system is left alone (in the compressed state) for some time, the local
stress distribution at the interface will decrease (or relax,
because of thermal excitation over the barriers), while the area of real
contact simultaneously increases. This will result in an increasing
adhesion bond between the solids, and a decrease in the elastic deformation
energy stored in the solids: both effects will tend to
increase of the pull-off force. (Note: The elastic energy stored at 
the interface during the
compression phase is almost entirely given back during
slow pull-off.) Since we use a frequency
independent elastic modulus,
such time-dependent effects are, of course, not taken
into account in the analysis presented above.

\begin{figure}[htb]
\begin{center}
\begin{minipage}{\textwidth}
   \includegraphics[width=0.475\textwidth]{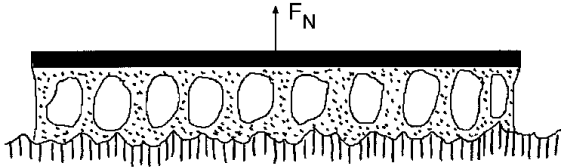}
\end{minipage} \\
\end{center}
\caption{
When the interaction between the ``glue'' film and the substrate is ``strong'',
the separation may involve internal rupture of the glue film rather than
detachment at the interface.
}
\label{PT4}
\end{figure}

The interfacial free energy is a sum of the adhesive part $U_{\rm ad}$,
which is proportional to the area of real contact, and the elastic
energy $U_{\rm el}$ stored in the strain field at the interface.
As long as $\Delta U = U_{\rm ad}+U_{\rm el} < 0$, a finite
pull-off force will be necessary in order to separate the
bodies. When the amplitude of the surface roughness increases, $\Delta U$
will in general increase and when it reaches zero,
the pull-off
force vanish.
Suppose now that an elastic slab
has been formed between two
solids from a liquid ``glue layer'',
which has transformed to the solid state after
some hardening time. For example, many glues consist of polymers
which originally are liquid, and slowly harden, e.g., via the
formation of cross bridges. In this case, if the original liquid
wets the solid surfaces, it
may penetrate into all surface irregularities and make
intimate contact with the
solid walls, and only thereafter harden to the solid state. Ideally,
this will result in a solid elastic
slab in perfect contact with the solid walls, and {\it without any interfacial
elastic energy stored in the system}, i.e., with $U_{\rm el} = 0$.
(In practice, shrinkage stresses may develop in the glue layer, which will
lower the
strength of the adhesive joint.)
Thus the last term in the expression for $\Delta \gamma_{\rm eff}$
vanish, and $\Delta \gamma_{\rm eff}$ will increase with increasing
surface roughness in proportion to the surface area.
This will result in an increase in the
pull-off force, but finally the bond-breaking may occur inside the
glue film itself\cite{[9]}, rather than at the interface
between the glue film and the solid walls (see Fig.\ \ref{PT4}); from here on
no strengthening of the adhesive bond will result from
further roughening of the confining solid walls.

Thus, the fundamental advantage of using liquid-like glues (which harden
after some solidification time), compared to pressure-sensitive adhesives
which consist of thin solid elastic ($E\approx 10^4-10^5 \ {\rm Pa}$)
films, and which develop tack only when squeezed between the solid surfaces,
is that in the former case no elastic
deformation energy is stored at the interface (which would be given
back during the removal process and hence
reduce the strength of the adhesive bond),
while this may be the case for the latter type of adhesive, unless the
interfacial stress distribution is able to relax towards the stress
free state (which requires the absence of cross links, or such a low
concentration of cross links that ``thick'' liquid-like polymer layers occur
at the interfaces).

If we define
$$\alpha = (q_0h_0)^2g(H)/2,\eqno(15)$$
$$\theta ={ E h_0^2 q_0 \over 4(1-\nu^2) \Delta \gamma},\eqno(16)$$
then (11) takes the form
$$\Delta \gamma_{\rm eff} =
\Delta \gamma \left ( 1 +\alpha - \theta f(H) \right )\eqno(17)$$
In what follows we will assume $\alpha <<1$ and neglect the $\alpha$ term
in (17). Note that without a low-distance cut-off
(i.e., $q_1/q_0 = \infty$), $f(H)=\infty$ for
$H \leq 1/2$ and it is clear that in this
limiting case no adhesive interaction will occur
{\it independent} of the magnitude of $\Delta \gamma$.
(This statement is only strictly true
as long as the attractive interaction responsible
for $\Delta \gamma$ is assumed to have
zero spatial extent.)
The physical reason is that in this case the
elastic energy stored in the deformation
fields in the solids will always be larger
than the adhesion energy which is proportional
to $\Delta \gamma$.
Note that for the important case $H\approx 1/2$, and if $\alpha << 1$,
(17) gives
$$\Delta \gamma_{\rm eff} \approx
\Delta \gamma \left [ 1-{1 \over 2} \theta \ {\rm ln}
\left ( {q_1\over q_0} \right )\right ]\eqno(18)$$
which (for $q_1/q_0 >> 1$)
is rather insensitive to the actual magnitude of $q_1/q_0$.

In the study above we have compared the free energies for the
case of complete contact between the rubber and the substrate,
with the case when no contact occur.
In reality, for large enough surface roughness
the free energy may be minimal for partial contact. Indeed, the
experimental results of Fuller and Tabor\cite{[3]}
suggest this to be the case (see Sec. 4), and in Sec. 5 we will consider
this case in greater detail.
\vskip 0.5cm

{\bf 4. Contact mechanics with adhesion: complete contact}

\begin{figure}[htb]
\begin{center}
\begin{minipage}{\textwidth}
   \includegraphics[width=0.475\textwidth]{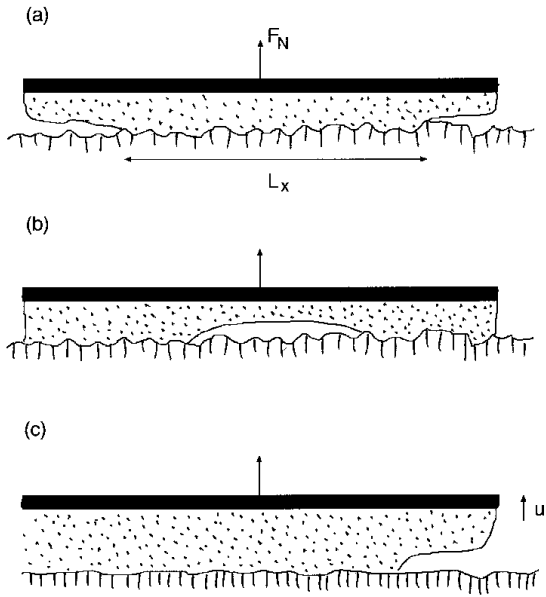}
\end{minipage} \\
\end{center}
\caption{
The block-substrate bond is broken by a crack propagating (a)
from the periphery of the contact area, or (b)
by a crack which has nucleated somewhere in the contact area, e.g., at
an imperfection. (c) Definition of the displacement $u$.
 }
\label{PT5}
\end{figure}

We consider the simplest possible case,
namely a rectangular elastic block
with flat surfaces, in contact with a
nominally flat substrate surface.
Assume
that the block has a height $L_z=L$ and
the bottom surface area $A_0=L_xL_y$.
Assume that the upper surface of the block is camped in the perpendicular
direction
[indicated by the thin (rigid) black slab in Fig. 5],
and pulled vertically with the force $F_{\rm N}$.
We assume that the bond between the block and the substrate
breaks via the propagation of an interfacial crack, which may nucleate
either (a) at the periphery of the contact area, or (b) at some point inside
the
contact area (see Fig. 5).
In the following we will make the simplifying assumption that
the stress in the block far away from the crack is uniaxial, as would be the
case if the
elastic film would be able to slide in the parallel direction. Thus, if the
upper clamped surface is moved upwards with the distance $u$, then the elastic
energy stored in the
block (in the absence of the crack) is $A_0L E(u/L)^2/2$.
Thus, assuming zero
surface roughness,
we write
the potential energy for the system as (see Fig.\  \ref{PT5})
$$U = -F_{\rm N} u +{1\over 2} A_0 L E \left ({u \over L}\right )^2
-A_0 \Delta \gamma $$
Minimizing this expression with respect to $u$ gives
$$F_{\rm N} = A_0 E u/L\eqno(19)$$
Now, consider $F_{\rm N}>0$. The block-substrate bond
clearly cannot break if the elastic energy stored in
the block is smaller than the surface
energy $A \Delta \gamma$ created when the block-substrate bond is broken.
We expect the bond between the block and the substrate
to break when the elastic energy becomes equal to the surface energy, i.e.,
$${1\over 2} A_0 L E \left ({u\over L}\right )^2
=A_0 \Delta \gamma $$
or
$$u =  \left ({2 \Delta \gamma L \over E}\right )^{1/2}$$
and the pull-off force $F_{\rm N}= F_{\rm c}$ [from (19)]:
$$F_{\rm c} =  A_0
\left ({2 \Delta \gamma E \over L}\right )^{1/2}\eqno(20)$$
The condition used above to determine the
adhesion force $F_{\rm c}$, namely that the
elastic energy stored in the block
equals the created surface energy, is only valid if the strain field
in the block is constant (which is the case in the
present simple geometry, but not in more complex geometries,
e.g., when a ball is squeezed against
a flat substrate). In general, this condition
must be replaced with the condition that
$U$ is stationary as the contact area
is varied, i.e., $\partial U /\partial A_0 = 0$.

The free energy minimization calculation
performed above can be extended to more complicated
systems. For example, when an elastic sphere (radius $R_0$) is in contact
with a substrate, the pull-off force becomes (see Appendix A)
$$F_{\rm c} = (3\pi/2) R_0 \Delta \gamma\eqno(21)$$
This result was first derived by Sperling \cite{[10]} and (independently) by
Johnson, Kendall and Roberts\cite{[4]}. Kendall has reported
similar results for
other geometries of interest\cite{[11]}.

Consider now the same problems as above, but assume that
the substrate surface has roughness described by the function
$z=h({\bf x})$.  We now study how the adhesion force is reduced
from the ideal value
(20) or (21) as the amplitude of the surface roughness is increased.
Let us first assume that the adhesive interaction is so
strong that the elastic solid is in contact
with the substrate everywhere. In this case we can still use the result (20),
but with $\Delta \gamma$ replaced by $\Delta \gamma_{\rm eff}$
as given by (11).
Thus if $\alpha << 1$ we get for a rectangular block
in contact with a nominally flat substrate:
$$F_{\rm c} = (F_{\rm c})_{\rm max} \left
[1-\theta f(H) \right ]^{1/2}\eqno(22)$$
where $(F_c)_{\rm max}$ is given by (20).
Similarly, for an elastic sphere in contact with a nominally flat substrate
$$F_{\rm c} = (F_{\rm c})_{\rm max} \left
[1-\theta f(H) \right ],\eqno(23)$$
where $(F_{\rm c})_{\rm max}$ is given by (21).
Note that $F_{\rm c} \rightarrow 0$ as $\theta f(H) \rightarrow 1$; when
$\theta f(H) = 1$ the elastic energy stored in the deformation field at the
interface equals the surface energy $\Delta \gamma A$ (where $A$ is the area of
real
contact),
and no ``external'' energy is necessary in order
to break the block-substrate bond.
When $\theta f(H) > 1$, the elastic energy stored at the interface
is larger than the gain in surface energy which would result from the
direct contact between the block and the substrate; this
state is stable only if the solids are squeezed against each other
with an external force.

\begin{figure}[htb]
\begin{center}
\begin{minipage}{\textwidth}
   \includegraphics[width=0.475\textwidth]{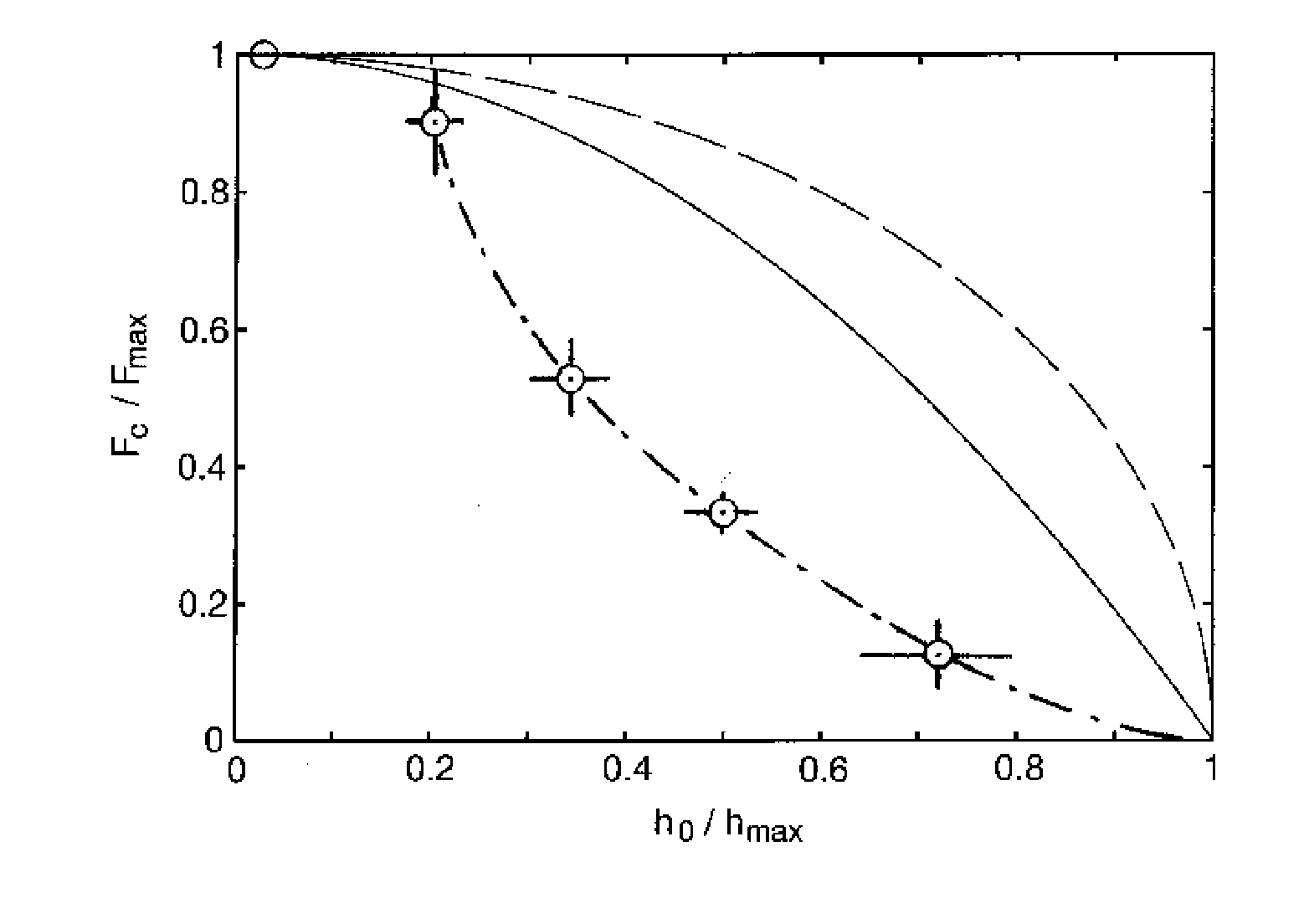}
\end{minipage} \\
\end{center}
\caption{
The pull-off force $F_{\rm c}$, in units of the maximum pull-off force,
as a function of the surface roughness amplitude $h_0$. The
solid and dashed lines are theoretical curves for a rectangular block,
and for a spherical ball, respectively, assuming complete contact in the
nominal contact area (see text).
The circles are experimental data from Ref. 3., and the dot-dashed line is
a guide to the eye  }
\label{PT6}
\end{figure}

\begin{figure}[htb]
\begin{center}
\begin{minipage}{\textwidth}
   \includegraphics[width=0.475\textwidth]{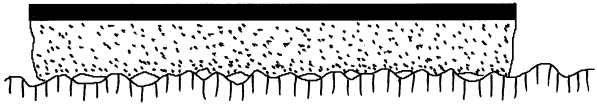}
\end{minipage} \\
\end{center}
\caption{
For ``large'' surface roughness
the free energy is minimal (when $F_{\rm N} = 0$) for {\it partial}
rubber-substrate
contact, rather than for complete contact.
 }
\label{PT7}
\end{figure}

In Fig.\ \ref{PT6} we compare the present theory with the experimental results
of
Fuller and Tabor for different surface roughness.  
The solid and dashed lines are theoretical curves for a rectangular block,
and for a spherical ball, respectively, assuming complete contact in the
nominal contact area.
The agreement between theory and experiment is good for small rms
roughness values, $h_0/h_{\rm max} < 0.2$ (where $h_{\rm max}$ is the
$h_0$-value for
which $\theta f(H)=1$, i.e., $h_{\rm max} = 2 [(1-\nu^2)\Delta \gamma /
Eq_0f(H)]^{1/2}$),
but for large $h_0$ the experimental
pull-off force falls somewhat below the theoretical prediction.
This may be due to the fact that
for ``large'' surface roughness
the free energy is minimal (when $F_{\rm N} = 0$) for {\em partial}
rubber-substrate
contact, rather than for complete contact (or zero contact),
as assumed above, see Fig.\ \ref{PT7}.

\begin{figure}[htb]
\begin{center}
\begin{minipage}{\textwidth}
   \includegraphics[width=0.475\textwidth]{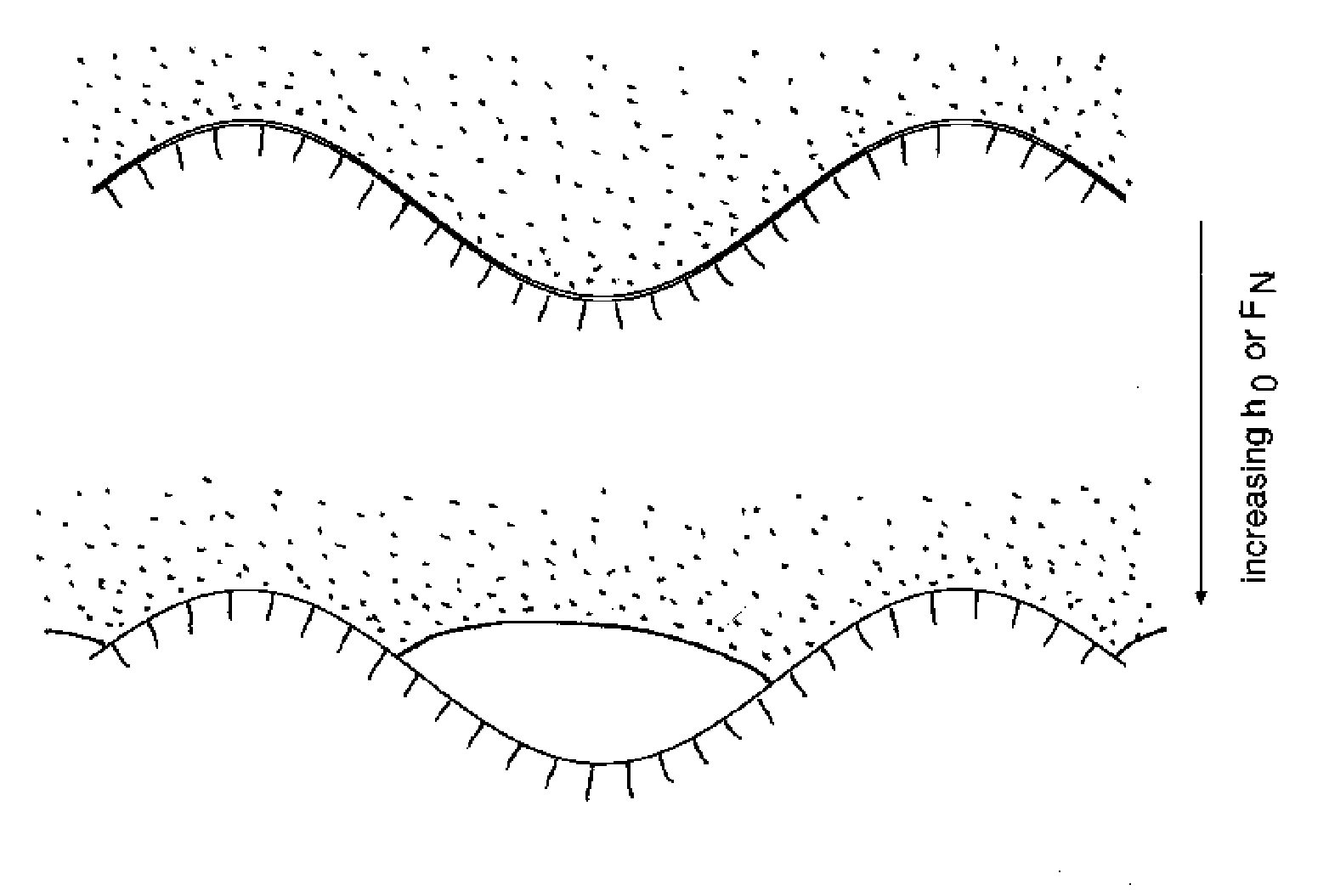}
\end{minipage} \\
\end{center}
\caption{
The detachment transition (schematic). For small surface roughness, 
complete contact occurs in the nominal contact area (top), while 
for large surface roughness there is a jump to partial contact (bottom).
 }
\label{PT8}
\end{figure}

\begin{figure}[htb]
\begin{center}
\begin{minipage}{\textwidth}
   \includegraphics[width=0.475\textwidth]{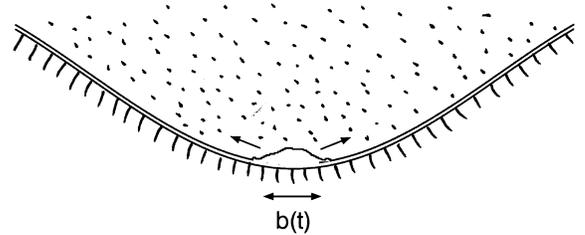}
\end{minipage} \\
\end{center}
\caption{ When the amplitude $h_0$ of the surface roughness, or the
pull-off force $F_{\rm N}$, is increased beyond a critical value,
a discontinuous detachment transition takes place from a state of
complete contact to partial
contact. The transition can be considered as resulting from the nucleation of a
crack at
the bottom of the valley, followed by rapid expansion of the crack
until it reaches a width of order $\sim \lambda /2$.
 }
\label{PT9}
\end{figure}

In fact, for surface roughness on a single length scale,
e.g., $z = h_0 {\rm cos}(q_0x)$, it is easy to convince oneselves
that there will be a
discontinuous {\em detachment transition} from complete contact to partial
contact (Fig.\ \ref{PT8}) when the pull-off
force (or the amplitude of the roughness $h_0$) is increased. This
can be seen directly if we consider a very narrow detached region at the
bottom of a valley as in Fig.\ \ref{PT9}. We can treat the detached region as
a crack of width $b$.  As is well known in that case \cite{[99]}
the stress at the crack edges will be proportional to
$(b/r)^{1/2}$, where $r$ is the distance away from a crack edge. Thus, the
local stress at a crack tip will increase with the width $b$ of
the crack, so that after nucleation the crack will expand to a finite size.
Thus partial detachment on a single length scale is a first order
transition. We have performed a preliminary study \cite{[12]}
[for a ${\rm cos}(q_0x)$-profile] which shows that
on increasing the pull-off force
(or increasing $h_0$ at vanishing external force)
the system first ``flips'' from a
state with complete contact, to another ``asperity contact''
state (Fig. 8) where the
width of the contact region is less than $\lambda /2$ as indicated in Fig. 8
(bottom).

Real surfaces do, of course,
exhibit roughness on many different length scales, and
the relation between the pull-off force and the center of mass
displacement is therefore likely to
be continuous for most systems of practical interest. Nevertheless,
during
pull-off rapid flip events may take part at the interface, where the solids
first undergoes local detachment in the valleys of the roughness profile,
followed at large enough pull off force by complete
detachment, the asperity contact areas detaching the last.
Because of the long-range nature of the elastic interaction, one may expect
a cooperative behavior of the detachment process, where detachment
in one local area may induce detachment in other interfacial surface areas.
Fuller and Roberts\cite{[8]} have studied the line of peeling (crack edge)
during pull off. For smooth surfaces the line is straight and peeling occurs
uniformly. Roughening the counterface makes the line increasingly irregular,
and peeling is intermittent, involving short sections of the
front at a time. This mode of behavior indicates variation in the
strength of the adhesion over the contact area as a result of the irregularly
fluctuating
surface roughness. The exact nature of the
detachment process and its possible collective behavior
represents an interesting problem for future studies.

Fuller and Tabor performed experiments with three different 
rubbers with very different elastic
modulus $E$. The dependence of the adhesion on the magnitude of $E$ is in good agreement with
the theoretical predictions above.

\vskip 0.5cm

{\bf 5. Contact mechanics with adhesion: partial contact}

We will now show that the discrepancy between theory and experiment for
$h_0 / h_{\rm max} > 0.2$ in Fig.\ \ref{PT6}
is due to rubber-substrate detachment, which
reduces the area of real contact and the pull-off
force for large surface roughness.
We assume again that the rough surface is a self affine fractal
with a long distance cut-off $\lambda_0 = 2\pi /q_0$. We will
refer to the ``asperities'' on the length scale $\lambda_0$ as
the macro asperities. The macro asperities are covered by
shorter wavelength roughness down to the lower cut-off length
$\lambda_1 = 2\pi /q_1$. We
assume the contact between the
rubber and the substrate to involve just a
fraction of the macro asperities. We will refer to a contact region
between a macro asperity and the substrate as the ``asperity contact area''.
We now make the basic assumption that the
rubber is in direct contact with the substrate in the asperity
contact areas and we will take into account the short-wavelength
surface roughness simply by using the
effective $\Delta \gamma_{\rm eff}$ introduced
above. Thus, the present problem reduces to the study of Fuller and Tabor,
except that we must replace $\Delta \gamma$ with $\Delta \gamma_{\rm eff}$.
Since $\Delta \gamma_{\rm eff} \rightarrow 0$ as $\theta f(H) \rightarrow
1$ it is still true that the pull-off force vanish
when $\theta f(H)=1$. However the pull off force before detachment
will not be the same.

\begin{figure}[htb]
\begin{center}
\begin{minipage}{\textwidth}
   \includegraphics[width=0.475\textwidth]{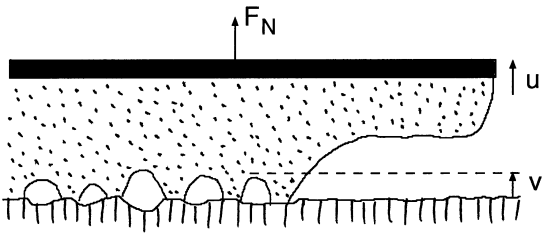}
\end{minipage} \\
\end{center}
\caption{
Definition of the displacements $u$ and $v$.
}
\label{PT10}
\end{figure}

\begin{figure}[htb]
\begin{center}
\begin{minipage}{\textwidth}
   \includegraphics[width=0.475\textwidth]{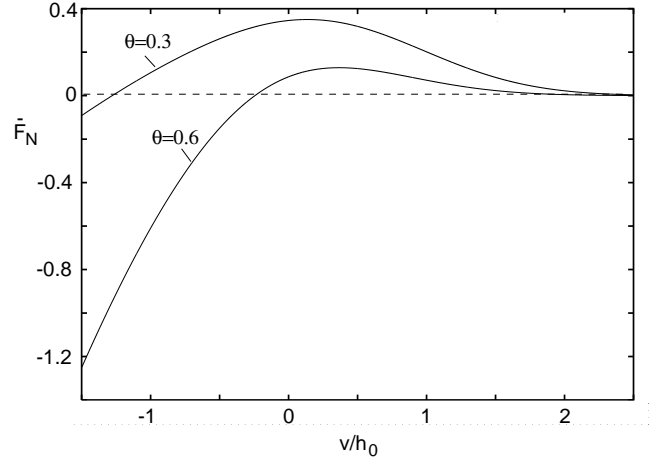}
\end{minipage} \\
\end{center}
\caption{
The normalized force $\bar F_{\rm N} = h_0 \ d \bar V (\theta, v/h_0)/dv$ as a
function of the
displacement $v$ (in units of $h_0$) of the bottom surface of the block.
For $\theta = 0.3$ and $0.6$, and with $f(H)=1$. }
\label{PT11}
\end{figure}

Let us consider the case of a
rectangular block in contact with a rough substrate. The
potential energy for the system is:
$$U = - F_{\rm N} u +{1\over 2} A_0 L E \left ({u-v\over L}\right )^2
+V(v),\eqno(24)$$
where $u$ and $v$ are the (lateral averaged) displacements of the upper and
lower
surface of the block (see Fig.\ \ref{PT10}),
and the block-substrate asperity interaction energy is
$$V=n_0A_0 \int_{z_{\rm c}}^\infty dz \ \phi(z) U_{\rm asp} (z-v), \eqno(25)$$
$n_0$ is the concentration of macro asperities, and $U_{\rm asp}$
the interaction energy between a substrate asperity and the elastic block,
and $z_{\rm c}$ is the smallest asperity height for which block-substrate
contact occurs.
The asperity height distribution $\phi (z)$ is assumed to be
Gaussian so that\cite{[13]}
$$\phi (z) = \left (\pi h_0^2\right )^{-1/2} {\rm e}^{-(z/h_0)^2}\eqno(26)$$
The radius $r$ of an asperity contact region can be related to the
compression $h=z-v$ via (during pull-off, $h < 0$)(see Appendix A and Ref. \cite{[4]}):
$$\bar h = \bar r^2 -(2\bar r)^{1/2}\eqno(27)$$
Here $ r = \alpha R \bar r$ and $h = \alpha^2 R \bar h$,
where $\alpha = (\pi \Delta \gamma_{\rm eff} /E^* R)^{1/3}$
(where $\Delta \gamma_{\rm eff} = \Delta \gamma [1-\theta f(H)]$),
defines
the dimensionless quantities $\bar r$ and $\bar h$.
The energy [see (A11)]:
$$U_{\rm asp} = E^* R^3 \alpha^5 \left ({8\over 15}\bar r^5
+\bar r^2 -{4\over 3} \bar r^3 \left (2\bar r\right )^{1/2}\right )\eqno(28)$$
Substituting (26) and (28) in (25) and defining $z=\alpha^2 R \bar z$ gives
$$V= n_0 A_0 \int_{\bar z_{\rm c}}^\infty d \bar z \ \alpha^2 R\left (\pi h_0^2
\right )^{-1/2} {\rm e}^{-\bar z^2 \left (\alpha^2 R/h_0 \right )^2}$$
$$\times E^* R^3 \alpha^5
\left ({8\over 15}\bar r^5
+\bar r^2 -{4\over 3} \bar r^3 \left (2\bar r\right )^{1/2}\right )\eqno(29)$$
Now, let us change integration variable, from $\bar z$ to $\bar r$.
Using $\bar z = \bar h +v/\alpha^2 R$ and (27) gives
$$d\bar z = d\bar r \left [2\bar r - (2\bar r)^{-1/2}\right ]$$
Thus,
$$V=
n_0A_0 \int_{\bar r_{\rm c}}^\infty d \bar r
\left [2\bar r - (2\bar r)^{-1/2}\right ]
\alpha^2 R\left (\pi h_0^2
\right )^{-1/2}$$
$$\times {\rm e}^{-\left [\bar r^2-(2\bar r)^{1/2}+v/\alpha^2R
\right ]^2 \left (\alpha^2 R/h_0 \right )^2}$$
$$\times E^* R^3 \alpha^5
\left ({8\over 15}\bar r^5
+\bar r^2 -{4\over 3} \bar r^3 \left (2\bar r\right )^{1/2}\right )\eqno(30)$$
where (see Appendix A) $\bar r_{\rm c} = (9/8)^{1/3}$.
Note that
$$ \Theta = {\alpha^2 R\over h_0} =
\left ({\pi \Delta \gamma_{\rm eff} R^{1/2} \over
E^* h_0^{3/2}}\right )^{2/3}$$
$$ \approx
\left ({\pi \over 4}\right )^{2/3} \theta^{-2/3}[1-\theta
f(H)]^{2/3}\eqno(31)$$
where we have assumed that
$1/R\approx q_0^2 h_0$. If we denote $\bar r = x$ for simplicity, then (30)
gives
$$V=
 -n_0A_0\Delta \gamma_{\rm eff} Rh_0 \bar V (\theta, v/h_0)\eqno(32a)$$
$$ \bar V = \surd \pi \Theta^2
\int_{(9/8)^{1/3}}^\infty d x
\left [2x - (2x)^{-1/2}\right ]$$
$$\times
{\rm e}^{- \left [\Theta ( x^2-(2x)^{1/2})+v/h_0
\right ]^2 }$$
$$\times
\left ({8\over 15}x^5
+x^2 -{4\over 3} x^3 \left (2x\right )^{1/2}\right )
\eqno(32b)$$
Minimizing (24) with respect to $u$ gives
$$F_{\rm N} = A_0 E (u-v)/L\eqno(33)$$
Similarly, minimization with respect to $v$ gives
$$A_0E{v-u\over L}+ {dV \over dv}=0$$
Using (33) this gives
$$F_{\rm N} =  {dV \over dv}\eqno(34)$$
Note that $F_{\rm N}$ only depend on $\theta$ and $v/h_0$.
In Fig. 11 we show $\bar F_{\rm N} = h_0 \ d \bar V (\theta, v/h_0)/dv$ as a
function
of $v/h_0$ for $\theta = 0.3$
and 0.6 [and with $f(H)=1$].
Fuller and Tabor\cite{[3]} determined the pull-off force from
curves such as in
Fig.\ \ref{PT11} by the condition $dF_{\rm N}/dv = 0$. However, this is usually
not the
correct condition: If the elastic energy in the block becomes equal to the
interfacial energy $A_0 \Delta \gamma_{\rm eff}$
before
the condition $dF_{\rm N}/dv = 0$ is
satisfied, then the pull-off force will be determined
by $U_{\rm el}=-U_{\rm ad}$.
The latter condition is relevant if the size of the block is large
enough (see below), which will be assumed to be the case in what follows.

\begin{figure}[htb]
\begin{center}
\begin{minipage}{\textwidth}
   \includegraphics[width=0.475\textwidth]{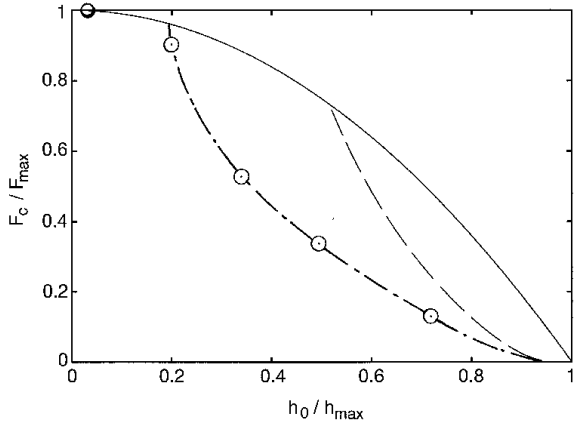}
\end{minipage} \\
\end{center}
\caption{
Solid line: the relation between the pull-off force and the
roughness amplitude, assuming complete contact between the block and the
substrate in the nominal contact area. Dashed line: The relation between
$F_{\rm c}$ and $h_0$ for partial contact
for $f(H)=1.0$. Points are the same experimental data of Fig. 6
 }
\label{PT12}
\end{figure}

The pull-off force is determined by the condition that the elastic
energy stored in the system is just large enough to break the
attractive block-substrate bond. This gives
$${1\over 2} A_0 L E \left ({u-v\over L}\right )^2
+V(v) = 0, $$
or, using (33),
$${1\over 2} A_0 L E \left ({F_{\rm c}\over A_0E}\right )^2
+V(v) = 0\eqno(35)$$
Using (32a)
this gives
$$F_{\rm c} \approx A_0 \left ({2\Delta
\gamma_{\rm eff} E \over L}\right )^{1/2}
{1\over 2 \pi} \left [\bar V (\theta,v/h_0)\right ]^{1/2},$$
or, comparing to (20),
$$F_{\rm c} \approx (F_{\rm c})_{\rm max} \left [1-\theta f(H)\right ]^{1/2}
\left [\bar V (\theta,v/h_0)\right ]^{1/2}/2\pi\eqno(36)$$
Note that $F_{\rm c} \sim L^{-1/2}$ so that
in the limit of large $L$,
$F_{\rm c}$ will be very small
and we can obtain the relevant $v/h_0$ to be used in
$\bar V (\theta, v/h_0)$ in (36) by putting
$F_{\rm N} = 0$ in (34), i.e., $dV/dv = 0$.
In Fig.\ \ref{PT12} (dashed line) we show the resulting
pull-off force as a function of $h_0$.
Note that there are no fitting parameters in the theory,
and that the calculation is in good qualitative agreement with 
the experimental trend, especially near  $h_{\rm max}$. In fact,
the present model calculation is only valid when the
asperity contact area is very small compared to $\lambda^2$
(only then is the JKR theory valid), i.e., the
theory holds strictly only for $h_0$ close to (but below) $h_{\rm max}$.
Thus, it is not surprising that the experimental reduction in the pull-off
force for $h_0$ well below $h_{\rm max}$ is somewhat larger 
than predicted by the theory.
Nonetheless, the overall qualitative form of detachment-induced
pull-off force reduction 
is in good agreement with
the
experimental data.

Let us close this section by discussing the two alternative pull-off conditions
(a) $dF_{\rm N} /dv = 0$ and (b) $U_{\rm el} = - U_{\rm ad}$ (or,
more generally, $\partial U_{\rm tot}/\partial A_0 = 0$). Condition (a)
correspond to
to a uniform (over the nominal contact area) detachment of the block-substrate
asperity
contact areas, while (b) correspond crack propagation, either
from the periphery of the nominal contact area, or from some point (crack
nucleation center)
inside the contact area. As stated earlier, if the block is big enough, case
(b) will
correspond to the smallest pull-off force, and will hence
prevail.

\vskip 0.5cm

{\bf 6. Discussion}

\begin{figure}[htb]
\begin{center}
\begin{minipage}{\textwidth}
   \includegraphics[width=0.475\textwidth]{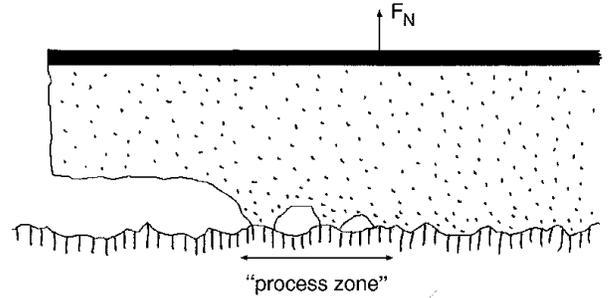}
\end{minipage} \\
\end{center}
\caption{
The transition from complete contact to detached area may involve a region of
partial
detachment, called the ``process zone''.}
\label{PT13}
\end{figure}

Consider an elastic block on a substrate. When the thickness $L=L_z$ of the
block
increases (but we assume $L_x >> L_z$ and $L_y >> L_z$),
the pull-off stress $F_{\rm c}/A_0$ decreases as $\sim L^{-1/2}$, see (20).
Thus, for large $L$ the (average) perpendicular stress at the block-substrate
interface will be very small (this is the reason for why glue films should be
very thin
in order to give a maximal pull-off force), and
the magnitude of the surface roughness alone
will determine whether the elastic media is in
complete contact with the substrate or only in partial contact.
(The same is true if instead of a block, an elastic ball is in contact
with the substrate. In this case the average stress in the contact area at
pull-off
decreases as $R_0^{-1/3}$, with increasing radius $R_0$ of the ball.) Of
course,
stress concentration will occur
at the crack tip,
so that partial detachment may occur
in a small region around the crack tip, even if complete contact occur far away
from the tip
inside the contact region,
see
Fig.\ \ref{PT13}. In the latter case, even if the crack propagate slowly, at
the crack tip
rapid flip events may occur as the individual block-substrate asperity contact
areas are broken.
This may lead to large energy dissipation, as the elastic energy stored
in the elongated bridges is lost during the rapid flip events,
and under those circumstances the pull-off force will be much larger than
predicted by (20) [or
(22)]. These rapid flips clearly
did not play any major role in the
experiments of Fuller and Tabor, but do occur in many practical
applications involving glues.
Usually the standard theory of crack motion can be used to treat these
more complicated cases, but $\Delta \gamma$ must now be replaced by the
strain energy release rate $G$, which is the energy needed
to propagate the crack by one unit area.
When only reversible processes occur at the crack tip (no rapid flip
processes),
$G=\Delta \gamma$ (or $\Delta \gamma_{\rm eff}$ for rough surfaces)
but if  cavity formation and fibrillar structures occur, $G$ may be
1000 times (or more) larger than $\Delta \gamma$.
The topic of designing glues exhibiting
large $G$ is of great practical importance.

The region in space where the block-substrate detachment occur at a crack edge
is usually
called the crack ``process zone'' (see Fig.\ \ref{PT13}). In some extreme cases
the width of the
this zone may become comparable to (or larger than) the width $L_x$ (or $L_y$)
of the
nominal contact region. In this case it is no longer correct (or useful) to
think about
the block-substrate bond breaking as involving crack propagation. This seems to
be the case for
many practical glues. The theoretical treatment of these cases cannot be based
on the theory
of crack motion,  but involves new physics,
such as the microscopic site of cavitation (i.e., the question
whether the nucleation occur right at the interface or in the bulk of the glue
film),
the concentration and spatial distribution of cavities, and the
evolution from cavities to fibrillar structures.
These processes have been intensively studied
recently for a flat probe geometry\cite{[16]}, where a 
block with a nominal flat
surface is squeezed
against a flat substrate covered by a thin (usually $L\approx 100  \ {\rm \mu
m}$) polymer
film acting as a pressure-sensitive-adhesive. After a short contact time the
block is removed
with a constant pull-off velocity, and the relation between the strain and
stress
is studies as function of time, while snap-shot pictures shows the geometrical
evolution of the adhesive film. It is found that very soft adhesive undergo
cavitation and fibrillation processes when subjected to a tensile stress.
A slight degree of cross-linking is 
beneficial for the stability of the fibrils, but excessive
cross-linking can lead to a premature failure of the fibrils, therefore
reducing
significantly the adhesion energy.

The voids first nucleate in the region which was last brought in contact with
the probe and thereafter
relatively homogeneously over the whole contact area. 
Nucleation will take place 
near the maxima in
the pull-off force. The cavities usually nucleate at the probe/film
interface.
The fact that nucleation occurs fairly homogeneously has been interpreted to
imply that the negative
hydrostatic pressure is fairly homogeneous under the probe surface. We do
believe this is indeed correct, but only after the nucleation of the cavities 
has started (see below).

Experiments with probe surfaces 
exhibiting different surface roughness 
have shown that even when cavitation and stringing occur,
the pull-off force
increased significantly when going from rough probe surfaces to
smooth ones\cite{[16]}. This is in accordance with the theory presented
earlier. Simultaneously, there appeared a striking difference 
in the morphology of the
de-bonding area. Thus, only the rough probe ($1.2 \ {\rm \mu m}$ rms roughness)
gave a significant fibrillar structure. The other probe surfaces 
($<0.1 \ {\rm\mu m}$ rms roughness)
did not evolve into a fibrillar structure so that, in the end, the adhesion
energy (the energy
to separate the probe from the substrate) where all quite comparable.

Let us discuss the process of cavity formation. Let us consider a thin
polymer film (thickness $L$) between two flat rigid surfaces. If the 
polymer is considered as
fully incompressible, then the pressure $p$ in the film is approximately\cite{[9],[16]}
$$p=p_{\rm ext}-E\epsilon \left ({r_0^2-r^2\over L^2}\right )\eqno(37)$$
where $p_{\rm ext}$ is the external pressure,
$\epsilon = \Delta L / L$ is the strain and
$r_0$ is the radius of the circular contact region.
The average pressure $\bar p = p_{\rm ext}-E\epsilon r_0^2/2L^2$.
It is interesting to note that this pressure distribution 
is similar to that for an
incompressible fluid (e.g., a polymer melt without cross links) (see, e.g.,
Ref. \cite{[1]}):
$$p=p_{\rm ext}-3\mu \dot \epsilon
\left ({r_0^2-r^2\over L^2}\right )\eqno(38)$$
where $\mu$ is the viscosity and
$\dot \epsilon = \dot L / L$. In fact, for a periodic oscillating strain,
$\dot \epsilon = - i \omega \epsilon$ and defining the complex elastic modulus
$E(\omega ) =
-i\omega \mu $, Eq. (38) takes the same form as (37) except for a factor of 3.
For a ``nearly''
incompressible material, say with the Poisson ratio $\nu = 0.49$, the pressure
distribution
becomes much flatter\cite{[16]}. However, the bulk modulus of polymers is of order
$10^{10} \ {\rm Pa}$,
while the elastic modulus $E \approx 10^4 \ {\rm Pa}$ (typical for pressure sensitive
adhesives at low
deformation rate) so that $0.5 -\nu \approx 10^{-6}$; under these conditions
the pressure
distribution in the polymer film will deviate
negligible from that calculated under the assumption
of an perfectly incompressible material. We must therefore ask why
the macroscopic cavities occur
uniformly in the contact area, in spite of the very non-uniform pressure
distribution
[Eq. (37)] which occur before the nucleation. We believe that 
the explanation of this puzzle may be related to detachment, as follows.

\begin{figure}[htb]
\begin{center}
\begin{minipage}{\textwidth}
   \includegraphics[width=0.475\textwidth]{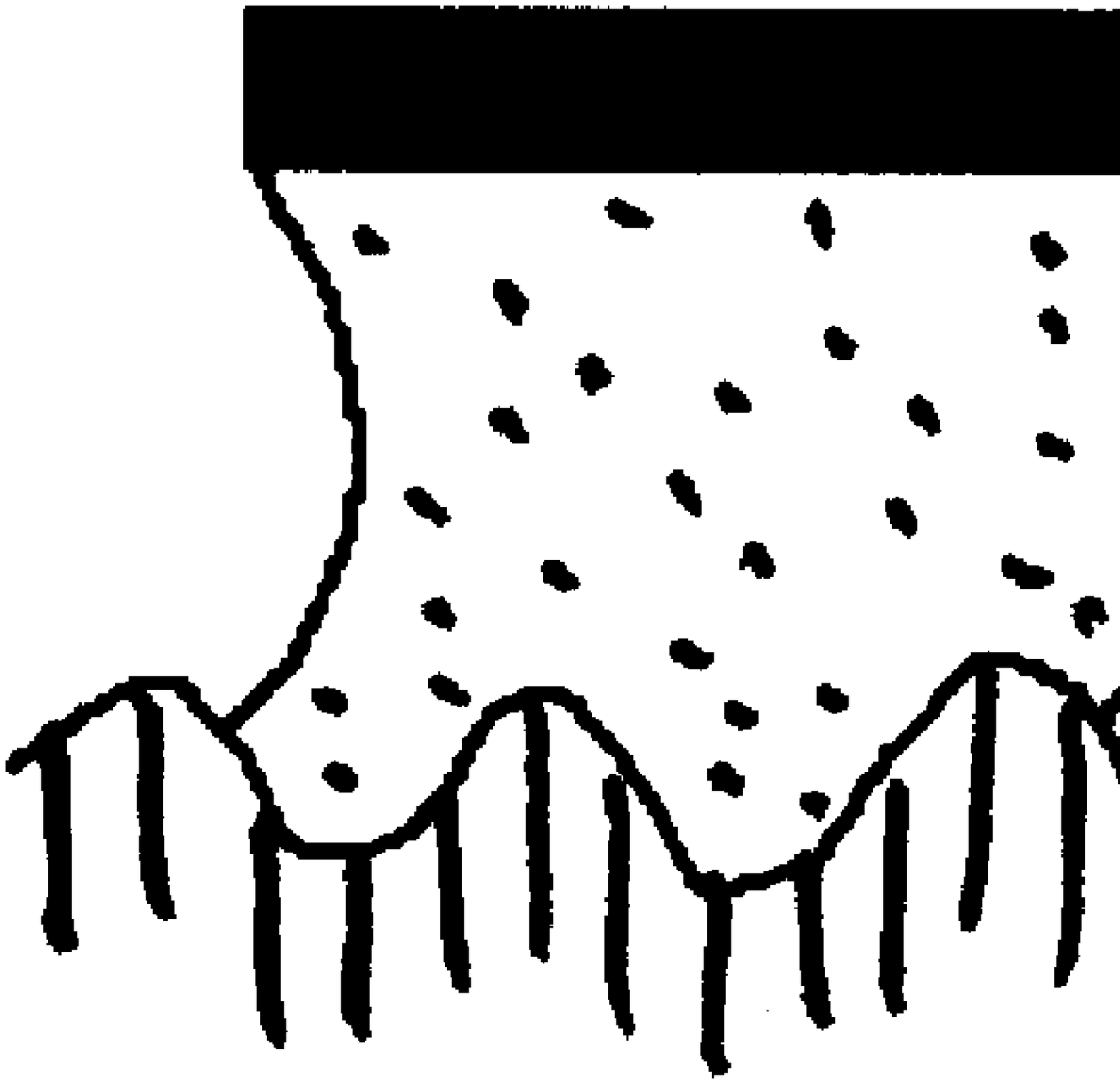}
\end{minipage} \\
\end{center}
\caption{
The external force $F_{\rm N}$ induce detached areas. The concentration of
detached areas is highest
in the center of the contact region, where the tensile stress would be highest
in the absence of the
detached areas. (Schematic.)}
\label{PT14}
\end{figure}

First, note that the typical maximal (average) pressure in a pull-off
experiment\cite{[16]} is of order
$0.4 \ {\rm MPa}$. Using (37) with $L= 100 \ {\rm \mu m}$,
$r_0 = 1 \ {\rm cm}$ (so that $r_0/L \approx 100$), and $E= 10^4-10^5 \ {\rm Pa}$
gives the true strain $\epsilon \approx 10^{-3}$ corresponding to the
displacement
$\Delta L = \epsilon L \approx 0.1 \ {\rm \mu m}$. Now,
the rms surface roughness of the probe surface was approximately $1 \ {\rm \mu
m}$.
Thus, it is clear that if a low concentration of microscopic
{\it local detachments occur at the interface when the
stress is increased} (see Fig. \ \ref{PT8}),
then this will locally reduce the stress in the contact region. If we assume
some characteristic
stress (``yield stress'') in order to induce a local detachment,
the detached areas will be distributed in such a way
(see Fig. \ \ref{PT14}) that a nearly uniform stress
may arise in the contact region even before any macroscopic detached regions
(cavities) can
be observed. As the strain is increased further, some of the microscopic
detached areas will grow into macroscopic cavities.
Hence, when the strain becomes so large (say 0.3) that (macroscopic)
cavities can be observed
it is clear that they must
be more or less uniformly distributed in the contact area.
This picture is consistent with the experimental observation that the
cavitation
stress is directly related to their shear modulus rather than their
bulk modulus\cite{[15]}.

\begin{figure}[htb]
\begin{center}
\begin{minipage}{\textwidth}
   \includegraphics[width=0.475\textwidth]{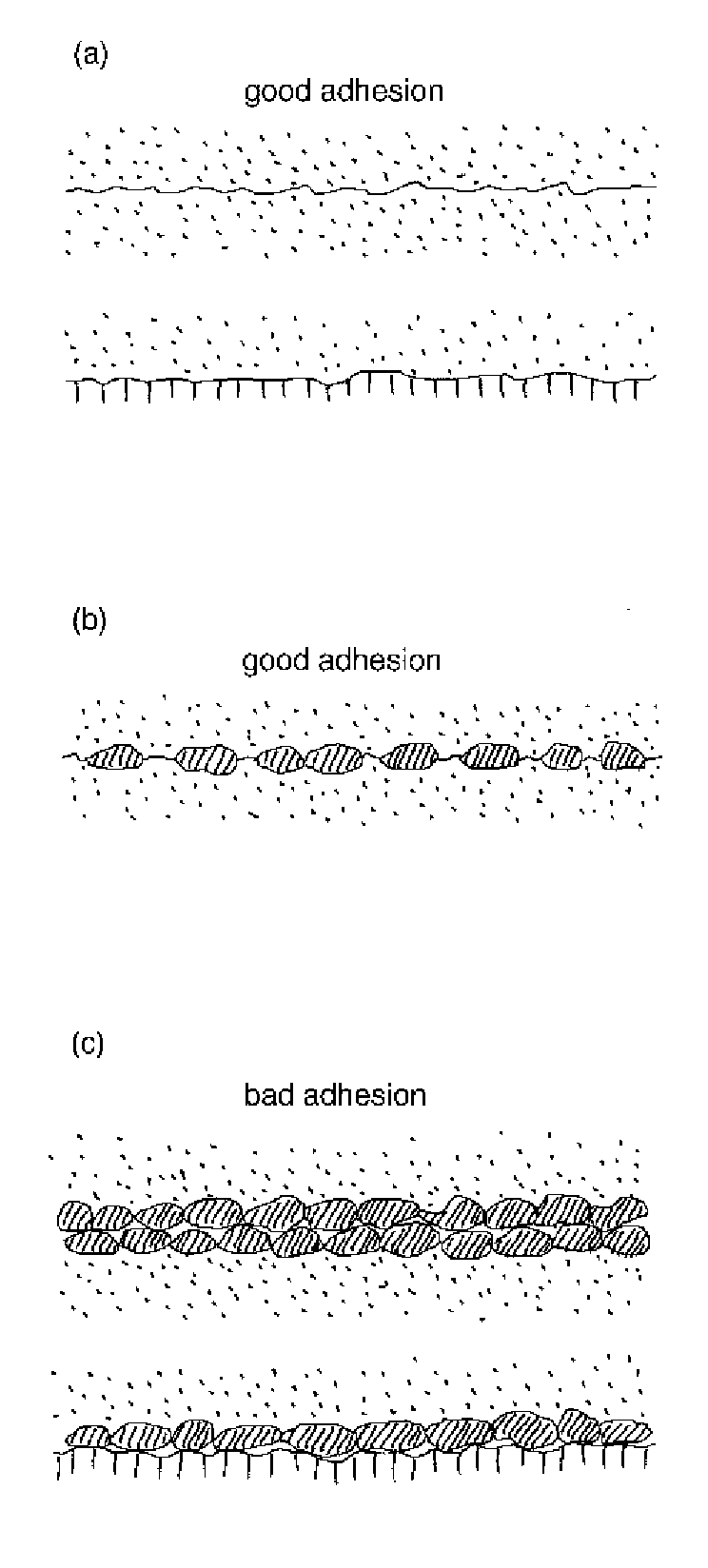}
\end{minipage} \\
\end{center}
\caption{ The influence of small particles (e.g., dust) on adhesion.
(a) The adhesion between two smooth, clean (identical) rubber surfaces,
or a rubber surface
and a smooth
hard substrate, is in general very good. (b) A monolayer (or less)
of small particles between two rubber surfaces may lead to a
pull-off force which is
even larger than for the clean rubber surfaces (see text). (c) A bilayer (or
more) of
particles between two rubber surfaces result in negligible adhesion. Similarly,
a monolayer of particles at a the interface between a hard solid and rubber
result in
negligible adhesion.}
\label{PT15}
\end{figure}

Finally, let us comment on the influence of
(small) contamination particles
(e.g., dust) on adhesion. It is generally believed that dusty rubber surfaces
provide bad adhesion. Now, while this is true in most practical
situations, one can imagine cases where it is not true. First, note that
the adhesion between two smooth, clean (identical) rubber surfaces
is in general very good (see Fig.\ \ref{PT15}) .
Now, if a monolayer (or less)
of small particles is deposited between the rubber surfaces, this
may lead to
an even larger pull-off force than for the clean rubber surfaces.
This follows from the fact that the particle-rubber adhesion may be stronger
than the
rubber-rubber adhesion [the van der Waals force is proportional 
to the polarizability, which is usually larger for hard (heavy) 
solids (e.g., rock)
than for rubbers]. However, if a bilayer (or more) of
particles occur between two rubber surfaces, negligible adhesion is observed,
as the separation now occur at the particle-particle interface. Similarly,
a monolayer of particles at a the interface between a hard solid and rubber
will result in
negligible adhesion.

\vskip 0.5cm

{\bf 7. Summary and conclusion}

We have studied the influence of
surface roughness on the adhesion of elastic solids.
Most real surfaces have roughness on many different length scales,
and this fact has been taken into account in our study. We have
considered in detail the case when the surface roughness
can be described by a self affine fractal, and shown that
when the fractal dimension $D_{\rm f}
>2.5$, the adhesion force may be strongly reduced.
We studied the behavior of the block-substrate pull-off force as a function
of roughness. For single scale roughness we find a partial
detachment transition before full detachment. Finally we studied
the full detachment transition for the self-affine fractal
surface, and found that total detachment
is characterized by exactly
the same parameter $\theta $ as in the simpler theory of
Fuller and Tabor. The partial detachment
which occur before full detachment however 
results in a very substantial reduction in the pull-off force
prior to full detachment. That is in good qualitative
agreement with experimental data.

\vskip 0.5cm

{\bf Acknowledgments}
B.P acknowledge a research and development grant
from Pirelli Pneumatici.
He also thanks BMBF for a grant related to
the German-Israeli Project Cooperation ``Novel Tribological Strategies
from the Nano-to Meso-Scales'', the EC for a ``Smart QuasiCrystals'' grant
under the EC Program ``Promoting Competitive and Sustainable GROWTH''.
He also thank SISSA for the warm hospitality during one month visit 
where part of this work was performed.
Work at SISSA was partly sponsored through
MURST COFIN, INFM,
the European Contract ERBFMRXCT970155 (FULPROP) and by INFM, PRA NANORUB.

\newpage

{\bf Appendix A}

\begin{figure}[htb]
\begin{center}
\begin{minipage}{\textwidth}
   \includegraphics[width=0.475\textwidth]{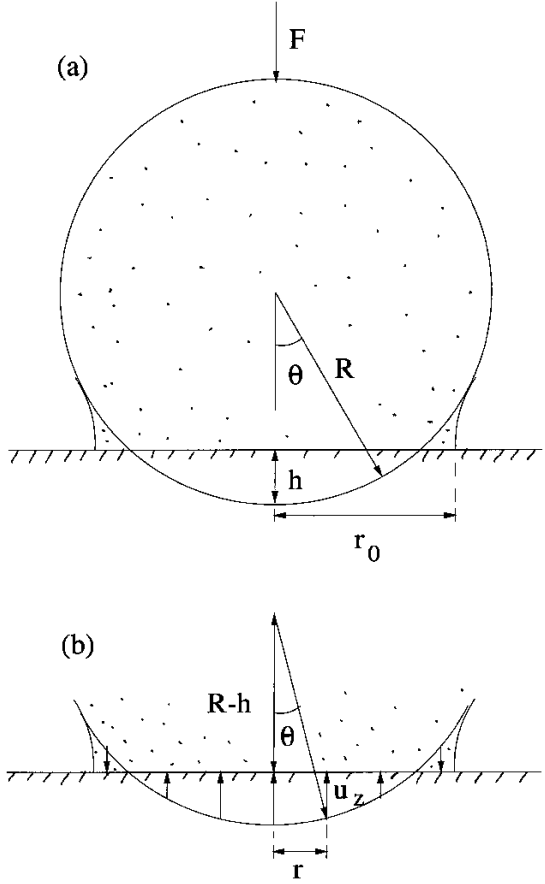}
\end{minipage} \\
\end{center}
\caption{
A rubber ball squeezed against a flat rigid substrate.}
\label{PT.Appendix}
\end{figure}

In this appendix we present, for the reader's convenience, a short
derivation of the JKR theory.
Consider an elastic sphere (radius $R$) in contact with a rigid flat solid
surface
(see Fig.\ \ref{PT.Appendix} ).
We assume that there is an attractive interaction between the two solids
so that the sphere deforms elastically at the interface forming 
a ``neck'' as indicated
in the Figure. Let $r_0$ be the radius of the (circular) contact area and
assume
that $h<<R$, where $R-h$ is the separation between the center of the sphere and
the substrate (see Fig.\ \ref{PT.Appendix}). In order for the deformed elastic
sphere to take the shape
indicated in Fig.\ \ref{PT.Appendix},
the surface of the sphere must displace as indicated by the arrows in
Fig.\ \ref{PT.Appendix} and given by the relation
$$u_z= h-R(1-{\rm cos} \ \theta)$$
But since $R \ {\rm sin} \ \theta = r$ we get
$${\rm cos} \ \theta = \left [1-(r/R)^2\right ]^{1/2}
\approx 1-r^2/2R^2$$
and thus
$$u_z \approx h\left (1-{r^2 \over 2hR}\right )\eqno(A1)$$
which is valid for $0<r<r_0$. Let us now determine
the pressure distribution which gives rise to the displacement (A1).
Since $h <<R$ (and $r_0 <<R$) we can determine the pressure distribution under
the
assumption that the surface of the sphere is locally flat.
Using the theory of elasticity, it has been
shown that when the surface of a semi-infinite elastic solid is exposed to the
pressure
$$\sigma =\sigma_0\left (1- {r^2\over r_0^2}\right )^{-1/2}+
\sigma_1\left (1-{r^2\over r_0^2}\right )^{1/2}\eqno(A2)$$
for $r<r_0$, and zero otherwise, then
the elastic deformation field (for $r<r_0$) becomes (see, e.g., Ref.
\cite{[14]}):
$$u_z = {\pi r_0 \over E^*}
\left [ \sigma_0+{1\over 2}\sigma_1 \left (1-{r^2\over 2 r_0^2}\right )
\right ]\eqno(A3)$$
where $E^*=E/\left (1-\nu^2\right )$.
Comparing (A3) with (A1) gives
$$\sigma_0 = {E^*\over \pi } \left ({h\over r_0}-{r_0\over R}\right
),\eqno(A4)$$
$$\sigma_1 = {E^*\over \pi } {2r_0\over R}.\eqno(A5)$$
Let us calculate the elastic energy stored
in the deformation field in the elastic
sphere in the vicinity of the substrate. This can be obtained using
the general formula
$$U_{\rm el} = {1\over 2} \int d^2x \ \sigma ({\bf x}) u_z({\bf x})\eqno(A6)$$
where the integral is over the surface area $r<r_0$.
Substituting (A2) and (A3) in (A6) gives
$$U_{\rm el} = \pi h  \int_0^{r_0} dr \ r
\left [\sigma_0\left (1- {r^2\over r_0^2}\right )^{-1/2}+
\sigma_1\left (1-{r^2\over r_0^2}\right )^{1/2}\right ]$$
$$\times \left (1-{r^2 \over 2hR}\right )$$
If we introduce
$\xi = 1-r^2/r_0^2$ we get
$$U_{\rm el} = {\pi h r_0^2 \over 2}  \int_0^1 d\xi
\left (\sigma_0\xi^{-1/2}+
\sigma_1\xi^{1/2}\right )
\left [1-{r_0^2 \over 2hR}(1-\xi) \right ]$$
$$={\pi h r_0^2\over 2}\left [ \left (2-{r_0^2\over hR}\right )\left
(\sigma_0+{\sigma_1\over 3}\right )
+{r_0^2\over hR} \left ({\sigma_0\over 3}+{\sigma_1\over 5} \right )\right
]\eqno(A7)$$
Substituting (A4) and (A5) in (A7) gives after some simplifications
$$U_{\rm el} = E^*
\left (h^2r_0-{2\over 3} {hr_0^3\over R}+{1\over 5}{r_0^5\over R^2}
\right )$$
In order to determine the radius $r_0$ of the contact area, we
must minimize the total energy under the
constraint that the $h= {\rm const.}$.
The total energy is given by the elastic energy plus the change in the surface
energy,
$-\Delta \gamma \pi r_0^2$, so that
$$U_{\rm tot} =
E^*
\left (h^2r_0-{2\over 3} {hr_0^3\over R}+{1\over 5}{r_0^5\over R^2}
\right )-\Delta \gamma \pi r_0^2$$
Let us introduce dimensionless
variables. If we define $\alpha   =(\pi \Delta \gamma /E^*R)^{1/3}$ and
introduce
$r_0 = \alpha R \bar r_0$ and $h = \alpha^2 R\bar h$ then the total energy
takes the form
$$U_{\rm tot} = E^*R^3\alpha^5\left (\bar h^2 \bar r_0-{2\over 3} \bar h \bar
r_0^3+
{1\over 5} \bar r_0^5 -\bar r_0^2 \right )\eqno(A8)$$
The force $F$ is given by
$$F= -{\partial U_{\rm tot} \over \partial h} =
 -{1\over \alpha^2 R} {\partial U_{\rm tot} \over \partial \bar h}$$
$$ =
E^* R^2 \alpha^3 \left (2\bar h \bar r_0 -{2\over 3} \bar r_0^3\right
)\eqno(A9)$$
The condition $\partial U_{\rm tot} /\partial \bar r_0=0$ takes the form
$$\left (\bar h -\bar r_0^2\right )^2 =2\bar r_0$$
with the solutions
$$\bar h = \bar r_0^2 \pm \left (2\bar r_0\right )^{1/2}\eqno(A10)$$
The two $\pm$-solutions correspond to different total energies, and the
correct solution is the one which minimize the total energy.
Substituting (A10) in (A8) gives
$$U_{\rm tot} = E^*R^3\alpha^5 \left ({8\over 15} \bar r_0^5 +\bar r_0^2\pm
{4\over 3}
\bar r_0^3 \left (2\bar r_0\right )^{1/2}\right ).\eqno(A11)$$
Thus the minus sign solution
gives the lowest energy.
The asperity snap-off is determined by the condition $dF/dr_0 = 0$. Using (A9)
and (A10) this
gives $\bar r_0 = \bar r_{\rm c} = (9/8)^{1/3}$
and the pull-off force
$F=-(3\pi /2) R \Delta \gamma$.

\end{multicols}
\end{document}